\documentstyle[aasms4,epsf]{article}

\def\deg{$^{\circ}$}
\def\bs{$\Box ^{\prime\prime}$}
\def\as{$^{\prime\prime}$} 
 
\def\bdm{\begin{displaymath}} 
\def\edm{\end{displaymath}} 
\def\beq{\begin{equation}} 
\def\eeq{\end{equation}} 
\def\bit{\begin{itemize}} 
\def\eit{\end{itemize}} 
\def\ben{\begin{enumerate}} 
\def\een{\end{enumerate}} 
\def\bfi{\begin{figure}[htb]} 
\def\bpfi{\begin{figure}[p]}
\def\mm{$\rm \mu m$} 

\def\llyc{$\rm L_{Lyc}$}

\def\lsol{$\rm L_{\odot}$}

\def\ea{{\it et al.}}
\def\brg{$\rm Br\gamma$}

\lefthead{B\"oker and Krabbe}
\righthead{NGC~253}

\begin{document}

\title{Mid-infrared [NeII] line emission from the nucleus of NGC~253}

\received{Jan. 14, 1998}
\accepted{March 6, 1998}
%\journalid{}{}
%\articleid{}{}

\author{Torsten B\"oker} 
\affil{Space Telescope Science Institute, 3700 San Martin Drive, 
Baltimore, MD 21218, U.S.A.}
\authoremail{boeker@stsci.edu}
\author{Alfred Krabbe} 
\affil{Max-Planck-Institut f\"ur extraterrestrische Physik, Garching, Germany}
\affil{DLR, Institute of Space Sensor Technology, Rudower Chaussee 5,
Berlin, Germany}
\authoremail{krabbe@dlr.de}
\author{John W.V. Storey} 
\affil{University of New South Wales, School of Physics, Sydney, Australia}
\authoremail{jwvs@newt.phys.unsw.edu.au}

\begin{abstract}  
We report on mid-infrared (MIR) continuum and line emission mapping of
the nucleus of NGC~253. The data, with a resolution of 1\as.4, reveal a 
double-peaked arc-like [NeII] emission region. Comparison with published data 
shows that the [NeII] arc is centered on the nucleus of the galaxy. 
The brightest [NeII] source coincides with the infrared continuum
peak. The interpretation of these results is complicated 
by the edge-on orientation of NGC~253, but a self-consistent explanation is 
starformation triggered by dynamical resonances in a barred potential.
\end{abstract}
%%%%%%%%%%%%%%%%%%%%%%%%%%%%%%%%%%%%%%%%%%%%%%%%%%%%%%%%%%%%%%%%%%%%%
%%%%%%%%%%%%%%%%%%%%%%%%%%%%%%%%%%%%%%%%%%%%%%%%%%%%%%%%%%%%%%%%%%%%%
%%%%%%%%%%%%%%%%%%%%%%%%%%%%%%%%%%%%%%%%%%%%%%%%%%%%%%%%%%%%%%%%%%%%%
%%%%%%%%%%%%%%%%%%%%%%%%%%%%%%%%%%%%%%%%%%%%%%%%%%%%%%%%%%%%%%%%%%%%%
\section{Introduction}\label{intro}
The prototypical starburst galaxy NGC~253 is a highly inclined (i=78\deg ,
\cite{pen81}) disk galaxy at a distance of about 3~Mpc (\cite{tul88}). 
The view towards its inner region is heavily obscured, and
the exact location of its nucleus has been the matter of some debate.
The astrometry of the various infrared sources in the central 10\as\ has
been discussed in detail by Kalas \& Wynn-Williams (1994, hereafter KW94)
in the near-infrared and \cite{ket93} in the mid-infrared.
From their $\approx$1\as\ resolution H, K, L, and M observations, KW94 
conclude that the brightest source in all the maps (their peak~1) lies at 
R.A.(1950)=$00^h45^m5^s.62$, 
Decl.(1950)=$-25^{\circ}33^{\prime}40^{\prime\prime}.2$. This position is
in very good agreement
with that of the emission peak at MIR wavelengths as determined by 
Keto \ea\ (1993) at 10~\mm\ from astrometric methods and Pi\~na \ea\
(1992) from low-level contour fitting to published maps with lower resolution. 
It is safe to assume that
the location of peak~1 is independent of wavelength up to 20~\mm , enabling
us to adopt the above coordinates for the continuum peak in our maps.

Peak~1 is not connected to any of the strong radio point 
sources found at 2~cm by Turner \& Ho (1985, hereafter TH85) and 6~cm by 
\cite{ulv91} and \cite{ant88}. The radio
sources are highly aligned along the major axis of NGC~253 at P.A. 51\deg .
All of them are most likely either radio supernovae or HII regions, based on
their spectral index. The only exception is the source 
TH~2 in TH85, a powerful flat spectrum radio source that they proposed 
to be the nucleus of NGC~253. TH85 argue that because of its high
radio luminosity ($10^4$~\lsol ) and its unique location
at the center of a synchroton disk this source probably is a compact
synchroton source, similar to those observed in active galactic nuclei (AGNs).
It lies close to a secondary MIR peak (peak~2) $\approx$ 2\as.2 northeast 
of peak~1 (\cite{pin92}, \cite{ket93}). However, as we will show, the 
radio nucleus is most likely not identical with any characteristic infrared
source. This view was strongly supported
from 0.5\as\ resolution NIR color maps by \cite{sam94}.
Their maps were referenced to other observations by cross-correlation of 
low level contours and show that the radio nucleus TH~2 coincides
with an extinction maximum of $\rm A_V\geq 24$. In addition, KW94 show that the
exact position of peak~2 is wavelength dependent in the sense that for
J, H, K observations it lies $\approx$ 1\as\ further east than in M-band.

The physical nature behind the various infrared sources is uncertain.
For example, according to \cite{for93}, the emission of the \brg -line 
at 2.17~\mm\ is strongest at peak~1. On the other hand, KW94 
show that peak~1 has a low PAH-feature/continuum emission ratio. 
Since PAH emission is usually indicative of ongoing star formation, KW94 argue
against peak~1 being an intense starburst which would be the natural explanation
for the \brg\ emission. To investigate this matter further, we have 
observed NGC~253 at MIR wavelengths, both in continuum and line emission. 
The forbidden $\rm ^2P_{3/2}-^2P_{1/2}$ transition of singly ionized Neon 
at 12.81~\mm\ traces the photoionization regions of hot stars and thus carries 
similar information than e.g. \brg , but at a much lower extinction: 
$\rm A_{2.17\mu m} \approx 2.5\cdot A_{12.8\mu m}$ (\cite{lut96,gen98}). 

In the next section we will describe how the data were obtained.
We also discuss the special observing techniques in the MIR and their
implications on data reduction. We present the results of our
observations in Sec. \ref{results} and interpret them in Sec. \ref{discussion}.
%%%%%%%%%%%%%%%%%%%%%%%%%%%%%%%%%%%%%%%%%%%%%%%%%%%%%%%%%%%%%%%%%%%%%
%%%%%%%%%%%%%%%%%%%%%%%%%%%%%%%%%%%%%%%%%%%%%%%%%%%%%%%%%%%%%%%%%%%%%
\section{Observations and data reduction} \label{data}
Our observations were carried out with MANIAC, the
Mid- And Near-Infrared Array Camera (\cite{boe97a}) built at the 
Max-Planck-Institut f\"ur extraterrestrische Physik in Garching, Germany.
NGC~253 was observed during five nights at the ESO 2.2m telescope 
in LaSilla, Chile in October 1996. The average seeing was below 1\as , therefore
the data have a spatial resolution defined by the diffraction limit of about 
1\as.4.

We used three narrow band filters 
($\frac{\Delta \lambda}{\lambda} \approx 0.015$) centered on
the [NeII] line (12.81~\mm) and two continuum positions shortwards 
(12.62~\mm) and longwards (13.00~\mm). 
This is essential for accurate continuum subtraction,
because strong Silicate absorption features and
emission from polycyclic aromatic hydrocarbon (PAH) make it 
difficult or even impossible to determine the continuum emission 
underlying the [NeII]~line from N-band photometry.
The mean of the two continuum images should represent the emission
underlying the [NeII], its subtraction from the 12.81~\mm\ map then 
reveals the distribution of the [NeII] emission. The total on-source
integration time was about 9600~s on the [NeII] filter and 4800~s on each
of the two continuum filters. 
 
Due to the high thermal background of the atmosphere at wavelengths around 
10~\mm, rapid readout of the detector, chopping of the telescope
secondary, as well as frequent beam switching is essential. We typically used
a chop amplitude of 15\as\, so that the source remained on the MANIAC
detector all the time. The correct telescope movement for the beam 
switch was defined with high precision by alternate guiding on each of the 
two (chopped) images of a nearby guide star.

The sky signal through the filters allowed single frame 
integration times of 80~ms before saturating the detector. 
Frames taken immediately after the chop 
movement were discarded because of the settling time of the secondary. 
After 1~min, the data (two sums of about 375 frames for each chop position) 
were transferred to the MANIAC PC and the beam switch was performed. 
After subtraction of the two beams, the residual background emission in our 
data is typically reduced by a factor of
$10^5$, as can be determined from empty regions in our images.

During the course of the night one gathers many such exposures,
which have to be coadded. Because of refraction by the
atmosphere, the source slowly changes position on the detector as the zenith
distance changes. We correct for this drift by coadding a few exposures
both at the beginning and the end of the observations to increase the 
signal-to-noise ratio, and determining the centroids of the well-defined
emission peak. Their difference in position gives the drift over this timeline.  
Assuming it to be linear with time, we can estimate the 
drift per exposure and correct each before coadding them.

The chop amplitude and direction, required for the correct overlay of the 
three source images on the detector,
was accurately determined by observing a point source in
an identical way, and fitting Gaussians to its three images. 
Finally, the resulting maps were calibrated for (slightly)
different filter transmissions and a mean image for each filter 
was generated from the data of the individual nights. 
Because the 12.81\mm\ filter showed a ghost reflection during the October 1996
run, the filter maps had to be deconvolved with their point spread functions
(PSF) before subtraction. We used a CLEAN algorithm (\cite{hog74}) with 
observations of reference stars as the PSF for each filter. Since the
observations described here are diffraction limited, the PSF is much more 
stable and the deconvolution more reliable than is normally the case.

Flux calibration of our data was performed by assuming that the 12.8~\mm\ flux
density of the nearby star $\beta$ Ceti is 53~Jy. This number was extrapolated 
from the 12~\mm\ IRAS flux and a blackbody temperature of 4200~K. 
The flux density inside a 5\as .5 aperture at 12.8~\mm\ is 17~Jy which
should be compared to the value of 11.2~Jy given by \cite{rie75} in
the same aperture at 12.6~\mm . The total flux density in our map is
35~Jy, about a factor of 2.5 higher than the result
of Keto \ea\ (1993), measured at 12.5~\mm . On the other hand, the 
[NeII] flux we derive from the central
7\as .6 is $2\cdot 10^{-14}$~W/m$^2$, roughly 60\% of the result by 
\cite{roc91}. The data thus seem to agree to some extent, although it should be 
stressed that a direct comparison between various data sets is problematic
because of the differences in filter passbands and bandwidths. This is especially 
true in the MIR since this regime is rich in spectral features.
%%%%%%%%%%%%%%%%%%%%%%%%%%%%%%%%%%%%%%%%%%%%%%%%%%%%%%%%%%%%%%%%%%%%%
%%%%%%%%%%%%%%%%%%%%%%%%%%%%%%%%%%%%%%%%%%%%%%%%%%%%%%%%%%%%%%%%%%%%%
\section{Results} \label{results}
\subsection{Morphology of the continuum and [NeII] emission}
The results of the above data reduction are presented in Fig. \ref{fig1}
(Plate 1). We show in panel A) the ``raw'' 12.81~\mm\ filter map before 
deconvolution with the PSF and in B) the deconvolved 12.81~\mm\ map. Panels
C) and D) contain the two deconvolved continuum maps 
at 12.62~\mm\ and 13.00~\mm . Finally, panel E) shows the 
resulting [NeII] line map, calculated as described in Sec. \ref{data}.
The continuum maps show a good overall
agreement with other MIR maps at comparable resolution
(\cite{pin92,ket93}), although there are differences in the detailed
emission morphology. This emphasizes our point that a reliable
continuum subtraction is only possible with narrow band continuum filters
neighboring the emission line.

Since both the strongest two continuum sources and the elongated emission
structure along the major axis of the NGC~253 disk at P.A. 51\deg\ 
(\cite{pen81}) are easily identified, 
we can safely adopt the astrometry and terminology used by KW94 (Sec. \ref{intro}). 
For the remainder of this paper, we will identify our strongest continuum source
with peak~1 and the secondary continuum source with peak~2, keeping in mind that
at NIR wavelengths peak~2 is offset by $\approx$ 1\as\ to the east.

The most striking result of the [NeII] map is that 
the emission forms a double-peaked, arc-like structure. The brightest
source coincides with peak~1, adding to the evidence that it is indeed
powered by a burst of star formation. The opposite end of the arc seems to
coincide with peak~2, but is somewhat elongated in E-W direction. 
The [NeII] morphology is in excellent agreement with the \brg\ map of
\cite{for93}.

It is important that the [NeII] emission from the central $\approx$ 5\as\
is aligned with the major axis of the NGC~253 optical disk. 
In Fig. \ref{fig2} we show again contour plots of the [NeII] and continuum 
emission to demonstrate its relative orientation. 
The line of nodes at P.A. 51\deg\ is indicated by the dashed
line. The dotted line shows the orientation of the NIR bar detected first by
\cite{sco85}. The [NeII] morphology and its 
relation to the NIR bar will be further discussed in Sec. \ref{discussion}.
%%%%%%%%%%%%%%%%%%%%%%%%%%%%%%%%%%%%%%%%%%%%%%%%%%%%%%%%%%%%%%%%%%%%%
\subsection{The radio nucleus}
Adopting the coordinates of peak~1 of KW94, we can relate 
our maps to the astrometry at longer wavelengths. 
In Fig. \ref{fig2}, we have marked the
location of the compact radio sources of TH85 on both the [NeII] and
the continuum emission. As can be seen, the major axis of the [NeII] emission 
coincides well with the radio axis. The radio nucleus TH~2
- or extinction peak according to \cite{sam94} - 
does fall on none of the [NeII] or continuum peaks.
Rather, TH~2 lies inside the arc-like [NeII] emission region. 
This supports the view of TH85 that TH~2 is most likely not an HII region, but
a synchroton source in the dynamical center of NGC~253. All other radio point 
sources are distributed rather randomly in the plane of the [NeII] emission, 
supporting the idea of them being radio supernovae (TH85). 

To summarize, the location of the nucleus of NGC~253 lies in the center 
of the [NeII] emission arc, about 2\as.2 northeast of peak~1 in a region 
with high extinction. 
%%%%%%%%%%%%%%%%%%%%%%%%%%%%%%%%%%%%%%%%%%%%%%%%%%%%%%%%%%%%%%%%%%%%%
%%%%%%%%%%%%%%%%%%%%%%%%%%%%%%%%%%%%%%%%%%%%%%%%%%%%%%%%%%%%%%%%%%%%%
\section{Interpretation and discussion} \label{discussion}
%%%%%%%%%%%%%%%%%%%%%%%%%%%%%%%%%%%%%%%%%%%%%%%%%%%%%%%%%%%%%%%%%%%%%
\subsection{The bar in NGC~253}
From 10\as\ resolution K-band imaging, \cite{sco85} find a
prominent NIR bar at P.A. 68\deg . However, the NIR continuum of the 
central $\approx$ 10\as\ in all high resolution NIR maps (e.g. 
Forbes \ea\ 1993, KW94, Sams \ea\ 1994) is oriented at P.A. 51\deg ,
roughly parallel to the major axis of the optical disk. 
This change in orientation is also visible as an isophote twist in \cite{sco85} 
and all MIR maps (\cite{pin92,ket93}, Fig. \ref{fig2}).
What is the reason for this behavior?
From kinematical studies of the CS(2-1) line at 98~GHz, \cite{pen96}
have tried to answer this question. They derived a model for the dynamical 
processes in the center of NGC~253.
In short, they find five prominent CS emission spots. The two innermost spots
are aligned along the axis of the radio knots, i.e. the major axis of
NGC~253. They are separated by $\approx$ 3\as\ and positioned symmetrically 
on either side of the radio nucleus. Two other knots lie again symmetrically
to the nucleus, but are aligned with the large scale NIR bar. 
Their observations are well 
explained by a rotating gas torus around the dynamical center of NGC~253.
The torus is the response of the viscous molecular gas to the potential 
of the NIR stellar bar. Gas clouds in the inner $\approx$ 10\as\ move along 
elliptical $x_2$ orbits whose major axis is aligned 
perpendicular to the bar (Athanassoula 1992a,b). 
At the inclination of NGC~253, these orbits appear
to be oriented parallel to the major axis of the optical disk as
demonstrated in Fig.~5 of \cite{pen96}. The gas clouds outside of 
$\approx$ 10\as\ move on $x_1$ orbits, their major axis being
aligned with the bar. The existence of $x_2$ orbits depends on the presence of
at least one Inner Lindblad Resonance (ILR). For the case of NGC~253, 
\cite{arn95} have shown that there are in fact two ILRs, one at
a radius of 25\as, the other close to the nucleus. It is at the inner ILR, where
the orientation of the dominant orbits changes from $x_2$ to $x_1$ 
(e.g. \cite{tel88}). Because of
orbit crowding, massive star formation occurs predominantly at the apocenters 
of the $x_2$ orbits (\cite{ath92b}), oriented symmetrically on either
side of the nucleus. Therefore, \cite{pen96} interpret their 
CS results as evidence for a rotating torus of dense gas around the radio
nucleus TH~2. These results are in agreement with observations of other
tracers of high density gas like HCN (\cite{pag95}).
%%%%%%%%%%%%%%%%%%%%%%%%%%%%%%%%%%%%%%%%%%%%%%%%%%%%%%%%%%%%%%%%%%%%%
\subsection{The [NeII] arc - a starformation ring}
Based on their model, \cite{pen96} predict that higher resolution line 
mapping should prove that gas emission from the central 10\as\ (145~pc) of 
NGC~253 is aligned with the radio knots rather than with the optical disk.
In that context, our results provide an independent confirmation of this 
scenario. Firstly, the [NeII] emission is well aligned with the radio 
knots (Figs. \ref{fig1} (Plate 1) and \ref{fig2}). Secondly, based on the
structure of the [NeII] emission, we also
confirm the distribution of the star forming material in a ring around the 
radio nucleus. The ring has a rotation velocity
of $\approx$ 60~km/s as seen in the CS data by \cite{pen96}, the western
side moving away from the observer. Unfortunately, our
observations are unable to resolve the velocity structure of the [NeII],
this has to await higher resolution line mapping, e.g. with a cryogenic
Fabry-Perot interferometer.

Our data only show weak emission from the southeastern half of the
ring. On the other hand, the [FeII] map of \cite{for93} does show 
emission from this region (their source B).
Thus, we favor the interpretation
of the double-peaked morphology as a starburst ring with a diameter of 
$\approx$ 4\as\ (60~pc). The total observed flux from a circular aperture
with 1\as.4 (20~pc) diameter centered on peak~1
is $1.7\cdot 10^{-15}$~W/m$^2$. After dereddening with 
$\rm A_V=24$ (Sams \ea\ 1994) in a mixed case scenario, this corresponds 
to $7.7\cdot 10^5$\lsol\footnote{We have adopted 
$\rm A_{12.8\mu m}=0.04\cdot A_V$ (\cite{gen98})}.

With the conversion factor $\rm \frac{L_{Lyc}}{L_{[NeII]}} = 64$ (\cite{gen98}),
the intrinsic Lyman-continuum luminosity is $4.9\cdot 10^7$\lsol.
This value should be used with caution given its uncertainties,
but it compares well to other bright star forming regions
like the core of 30~Doradus in the LMC (\cite{bra96}) or the brightest 
SFR in the gaseous ring of IC~342 (\cite{boe97b}). This supports the view that
the [NeII] emission probably stems from individual giant molecular clouds
(GMCs) that actively form stars and are located in a ring at the ILR. 
We will extend the comparison to a similar structure in IC~342 in 
the next section.
%%%%%%%%%%%%%%%%%%%%%%%%%%%%%%%%%%%%%%%%%%%%%%%%%%%%%%%%%%%%%%%%%%%%%
\subsection{NGC~253 and IC~342 - two of the same kind?}
It is interesting to note the close similarities between NGC~253 and IC~342,
another nearby late type spiral. B\"oker \ea\ (1997b) describe
the dynamical processes in the central 100~pc of IC~342 as
derived from NIR integral field spectroscopy. Table \ref{tab1}
compares the two galaxies. 
\begin{table}
\dummytable \label{tab1}
\end{table}
The dynamical processes and the morphology of the molecular gas are almost 
identical. This was also noted by Paglione \ea\ (1995). One important 
difference, however, is that IC~342 does not house a central synchroton source. 
In fact, in the case of IC~342, there is no radio emission from the dynamical
center (\cite{tur83}). 
Rather, it is occupied by a cluster of young red supergiants that dominate
the NIR continuum. For NGC~253, there is no obvious NIR component at the
location of the radio nucleus, but the edge-on orientation and patchy extinction
might complicate its detection. Both the diameter and rotation velocity of
the molecular rings are very similar. This adds to the increasing evidence 
that stellar bars and small star formation rings on scales of 50-100~pc are
a common feature in late type spirals. This points to an evolutionary 
connection between spirals of different Hubble type, as suggested by
\cite{pfe94}.
%%%%%%%%%%%%%%%%%%%%%%%%%%%%%%%%%%%%%%%%%%%%%%%%%%%%%%%%%%%%%%%%%%%%%
%%%%%%%%%%%%%%%%%%%%%%%%%%%%%%%%%%%%%%%%%%%%%%%%%%%%%%%%%%%%%%%%%%%%%
\section{Summary}
We have reported on MIR line and continuum mapping of the central 200~pc
of NGC~253. The main results can be summarized as follows: \\
1) The continuum maps confirm that
the nucleus of NGC~253 is not identified with any strong IR source.\\
2) The [NeII] map reveals two peaks separated by about
4\as\ and connected by an arc-shaped emission region. \\
3) The nucleus lies inside the [NeII] arc. \\
4) We interpret the morphology of the [NeII] map as evidence for a bar-triggered
starformation ring with a diameter of 50~pc, similar to that seen in IC~342.

It is a pleasure to thank Thomas Lehmann for his invaluable
help before, during, and after the campaign. We are indebted to
the ESO 2.2m-team for their continuous support, and to
Rolf Chini and Craig Smith who helped a great deal with the observations.
We also would like to acknowledge Thomas Ott for his expertise on 
deconvolution algorithms, and the anonymous referee whose comments led
us to re-examine the point spread function and thereby substantially 
improve this paper. JWVS thanks the Alexander von 
Humboldt Stiftung for their support during the development of MANIAC, and the
Australian DIST for travel support.
%%%%%%%%%%%%%%%%%%%%%%%%%%%%%%%%%%%%%%%%%%%%%%%%%%%%%%%%%%%%%%%%%%%%%

%%%%%%%%%%%%%%%%%%%%%%%%%%%%%%%%%%%%%%%%%%%%%%%%%%%%%%%%%%%%%%%%%%%%%
%%%%%%%%%%%%%%%%%%%%%%%%%%%%%%%%%%%%%%%%%%%%%%%%%%%%%%%%%%%%%%%%%%%%%
\newpage
\figcaption[fig1.ps]{MIR maps of the central 15\as\ (200~pc) of NGC~253. 
A) ``raw'' 12.81\mm\ filter map before CLEANing. B) 12.81\mm\ filter map
after CLEANing. C) 12.62~\mm\ continuum map after CLEANing. 
D) 13.00~\mm\ continuum map after CLEANing.
E) [NeII] line map calculated as B - (C+D)/2. 
The resolution in all maps is 1\as.4
The coordinate origin is the continuum peak (peak~1) at 
R.A.(1950)=$00^h45^m5^s.62$, 
Decl.(1950)=$-25^{\circ}33^{\prime}40^{\prime\prime}.2$.
The contour levels for the three continuum maps are 
$4,8,12,16,20,30,40\cdots 100$\% of the peak continuum flux at 12.8\mm\
($1.6\cdot 10^{-16}$~W/m$^2$/\bs), and
$6,12,18,24,30,40,50\ldots 100$\% of the peak line flux 
(2.3~Jy/\bs) for the [NeII] map. \label{fig1}}

\figcaption[fig2.ps]{12.8~\mm\ continuum emission (top) and [NeII] line 
emission (bottom)
from the central 15\as\ (200~pc) of NGC~253. The coordinate origin is the continuum peak (peak~1) at 
R.A.(1950)=$00^h45^m5^s.62$, Decl.(1950)=$-25^{\circ}33^{\prime}40^{\prime\prime}.2$
(see text for discussion). The resolution in both maps is 1\as.4. Contour
levels are $4,8,12,16,20,30,40\cdots 100$\% of the peak flux
(2.3~Jy/\bs ) for the continuum and
$6,12,18,24,30,40,50\ldots 100$\% of the peak flux 
($1.3\cdot 10^{-15}$~W/m$^2$/\bs) for the [NeII] map. The location of the
non-thermal radio nucleus TH~2 (coincident with the NIR extinction peak of
Sams \ea\ 1994) as well as the thermal 2cm radio 
point sources of TH85 are marked by the indicated symbols for {\it Nuc}
and {\it SNR}, resp. The dashed line shows the orientation of 
$x_2$ orbits that form the starburst ring. Incidentally, this is identical 
with the major axis of the optical disk of NGC~253 at P.A. 51\deg . 
The dotted line denotes the orientation of the NIR bar at P.A. 68\deg .
\label{fig2}}

\newpage
\begin{deluxetable}{lcccl} \label{tab1}
\tablenum{1}
\tablewidth{0pt}
\tablecaption{NGC~253 and IC~342 - a comparison}
\tablehead{
\colhead{Property} & \colhead{IC~342} & \colhead{NGC~253} & \colhead{Ref.}}         
\startdata
Distance [Mpc] & 1.8 & 3 & 1,2 \nl 
Inclination & 25\deg\ & 78\deg\ & 3,4 \nl 
P.A. & 39\deg\ & 51\deg\ & 3,4 \nl
Morph. type & Scd & Sc & 5 \nl 
Radio core & no & yes & 6,7 \nl 
P.A. of NIR bar & 0\deg\ & 68\deg\ & 8,9 \nl 
No. of ILR's & 1 & 2 & 8,10 \nl 
Ring diameter [pc] & 70 & 60 & 8,11 \nl 
\llyc\ from brightest SRF [$10^7$ \lsol] & 1.7 & 4.9 & 8,12 \nl
\enddata
\tablecomments{References: 1) \cite{mcc89,mad92} 2) \cite{tul88} 3) \cite{new80}
4) \cite{pen81} 5) NASA extragalactic database (NED) 6) \cite{tur83} 
7) \cite{tur85} 8) \cite{boe97b} 9) Scoville \ea\ 1985 10) Arnaboldi \ea\ 1995 
11) Peng \ea\ 1996, \cite{pag95}, this work, 12) this work}
\end{deluxetable}

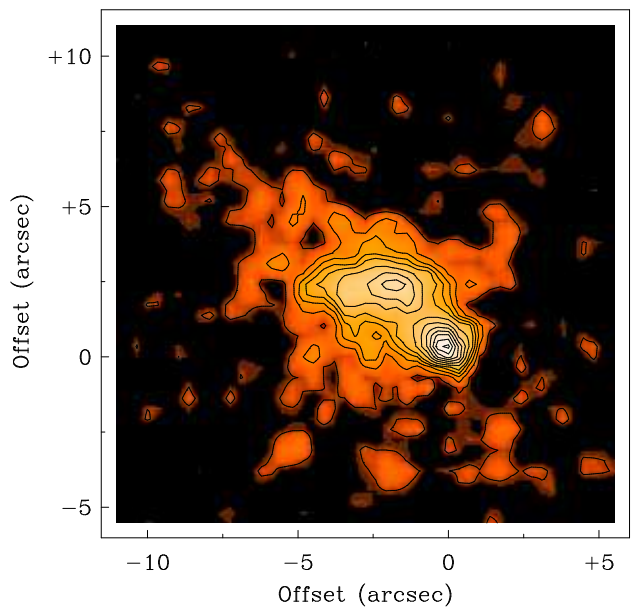
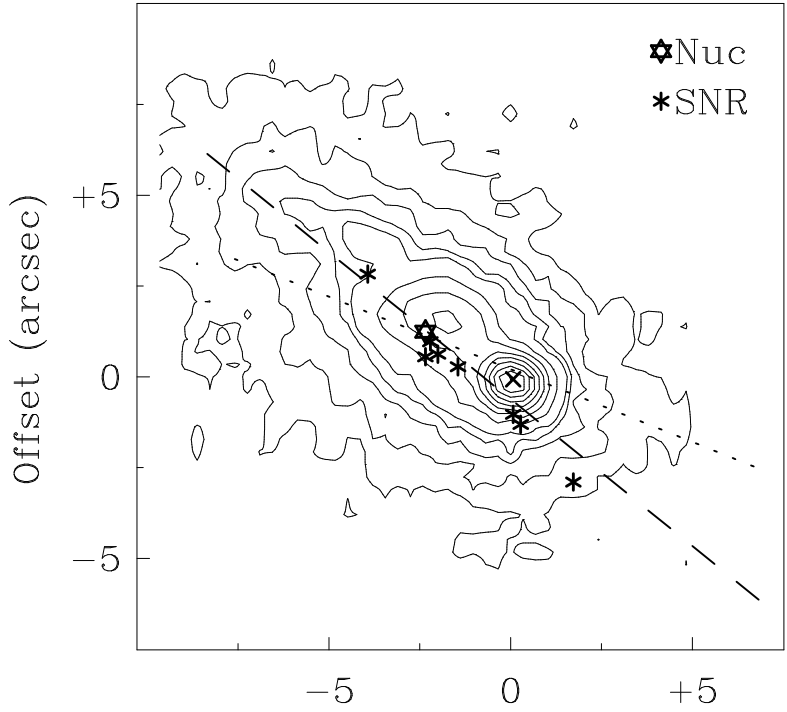
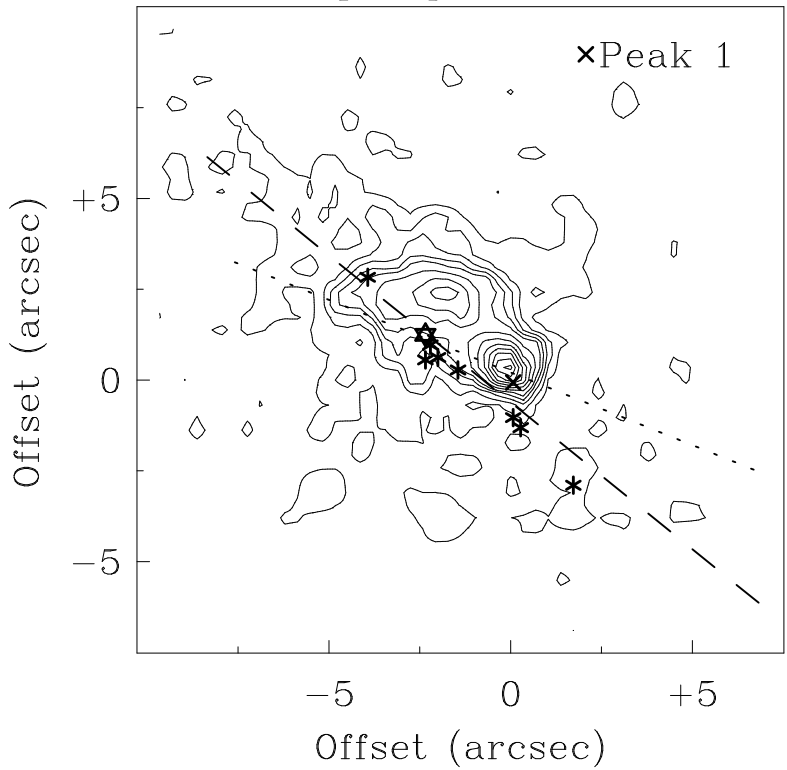
\end{document}